\documentclass[11pt]{article}
\usepackage{epsfig,latexsym}
\usepackage{amsmath}
\usepackage{tensor}
\usepackage{multibox}
\usepackage{verbatim}
\usepackage{mathrsfs}
\usepackage{amssymb}
\usepackage{epsf}

\DeclareMathAlphabet{\mathpzc}{OT1}{pzc}{m}{it}

 \csname
@addtoreset\endcsname{equation}{section}

\def\gz0{\gamma^{0}}

\def\scs#1{\section{\sc #1}}
\def\scss#1{\subsection{\sc #1}}

\def\a{\alpha}

\def\g{\gamma}

\def\e{\epsilon}

\def\h{\eta}

\def\x{\xi}

\def\vf{\varphi}

\def\be{\begin{equation}}
\def\ee{\end{equation}}
\def\bea{\begin{eqnarray}}
\def\eea{\end{eqnarray}}
\def\ba{\begin{array}}
\def\ea{\end{array}}
\def\bec{\begin{center}}
\def\ec{\end{center}}
\def\ba{\begin{align}}
\def\ena{\end{align}}

\def\12{\frac{1}{2}}

\def\pr{\partial}

\begin{document}

\begin{flushright}
{\today} \\
CPHT-RR078.0910 \\
LPT-ORSAY 10-65
\end{flushright}

\vspace{25pt}

\begin{center}


{\Large\sc On Climbing Scalars in String Theory}\\


\vspace{25pt}
{\sc E.~Dudas${}^{\; a,b}$, N.~Kitazawa${}^{\; c}$ and A.~Sagnotti$^{\; d}$}\\[15pt]

{${}^a$\sl\small Centre de Physique Th\'eorique,
\'Ecole Polyt\'echnique, CNRS\\
F-91128 Palaiseau \ FRANCE\\}e-mail: {\small \it
dudas@cpht.polytechnique.fr}\vspace{10pt}

{${}^b$\sl\small LPT, Bat. 210, Univ. Paris-Sud, F-91405 Orsay \
FRANCE}\vspace{10pt}

{${}^c$\sl\small Department of Physics, Tokyo Metropolitan University\\
Hachioji, Tokyo \\ 192-0397 JAPAN
\\ }e-mail: {\small \it
kitazawa@phys.metro-u.ac.jp}\vspace{10pt}

{${}^d$\sl\small
Scuola Normale Superiore and INFN\\
Piazza dei Cavalieri, 7\\I-56126 Pisa \ ITALY \\
e-mail: {\small \it sagnotti@sns.it}}\vspace{10pt}

\vspace{35pt} {\sc\large Abstract}\end{center}
\noindent {In string models with ``brane supersymmetry breaking'' exponential
potentials emerge at (closed--string) tree level but are not accompanied by
tachyons. Potentials of this type have long been a source of embarrassment in
flat space, but can have interesting implications for Cosmology. For instance,
in ten dimensions the logarithmic slope $\left| V^{\,\prime}/V\right|$ lies
precisely at a ``critical'' value where the Lucchin--Matarrese attractor
disappears while the scalar field is \emph{forced} to climb up the potential
when it emerges from the Big Bang. This type of behavior is in principle
perturbative in the string coupling, persists after compactification, could
have trapped scalar fields inside potential wells as a result of the
cosmological evolution and could have also injected the inflationary phase of
our Universe.}

\setcounter{page}{1}

\pagebreak




\scs{Introduction}\label{sec:intro}


Inflation is today a basic tenet for Cosmology \cite{cosmology}, while
supersymmetry breaking \cite{weinberg_ft} is perhaps the key issue that one
must face when trying to connect String Theory to the real world. It typically
destabilizes an original Minkowski vacuum, so that little can be done if one
insists on static backgrounds, aside from appealing to the Fischler--Susskind
mechanism \cite{fs} or to similar resummations around an original ``wrong
vacuum'' \cite{wv}. The two problems, however, can find a common ground in the
orientifolds \cite{orientifolds} where ``brane supersymmetry breaking''
\cite{sugimoto,bsb} occurs, whose vacua accommodate non--BPS combinations of
extended objects in such a way that supersymmetry, broken \emph{at the string
scale}, appears non--linearly realized in the low--energy supergravity
\cite{dm1}. Tadpoles do arise in these models, but to lowest order they are not
accompanied by tachyons, so that important information can be extracted taking
into account the exponential potentials that they bring about. On the other
hand, it is natural to wonder whether a supersymmetry breaking mechanism that
is intimately tied to the string scale could have played a role in the Early
Universe. With these motivations in mind, in this letter we elaborate upon a
phenomenon that, as we recently came to know, was actually discussed earlier,
from a different perspective, in \cite{townsend}\footnote{We thank E. Kiritsis
for calling to our attention refs.~\cite{townsend} after our results were
presented at ``String Phenomenology 2010'' \cite{augustopheno}. The paper of
J.~Russo in \cite{townsend} contains the classical solutions that we discuss in
Section \ref{sec:ten}, while a vast literature, that includes
\cite{attractor,halliwell,pli,exponential,bergshoeff}, is devoted to the
asymptotic behavior of similar models.}:~\emph{a scalar field can be compelled
to emerge from the Big Bang while climbing up an exponential potential}. This
happens when the logarithmic slope of the potential,
$\left|V^{\,\prime}/{V}\right|$, reaches a certain ``critical'' value, and
amusingly for the ten--dimensional Sugimoto model of \cite{sugimoto} it is
precisely critical.

There is an interesting analogy between this phenomenon and the motion in a
viscous medium of a Newtonian particle subject to a constant force $f$, for
which the equation of motion and its solution read
\be
m \ {\dot v}(t) \ + \ b \ v(t) \ = \ f \ , \qquad
v (t) \ = \ (v_0 - v_{l} ) \ e^{\,- \, \frac{b\,t}{m}} \ + \ v_{l} \ . \label{mechanical}
\ee
Insofar as $b$ is finite, there are two ``branches'' of solutions, depending on
whether the initial speed $v_0$ lies above or below the ``limiting speed''
$v_{l}= f/b$, while as $b\to 0$ the upper branch disappears altogether. In
String Theory the non--linear equations for a scalar field in the presence of
an exponential potential also admit two distinct branches of cosmological
solutions for logarithmic slopes that are small enough, and the climbing
solution is branch that in our mechanical analogy corresponds to $v_0 < v_{l}$.
When the logarithmic slope reaches a \emph{finite} critical value the other
branch ceases to exist, and as a result the scalar can only exit the Big Bang
while climbing up the potential. In the simplest one--field model of this type,
the transition occurs precisely when the Lucchin--Matarrese attractor
\cite{attractor} disappears.

With more scalar fields, and in four dimensions in particular, the dynamics
becomes far richer, but the phenomenon continues to occur, and in particular a
``critical'' logarithmic slope obtains again in the KKLT model \cite{kklt},
where the eventual descent is dominated by an attractor. These types of models
with potential wells actually afford another interesting possibility: climbing
scalars can get ``trapped'' in them as a result of the cosmological evolution.
This is seen very clearly in numerical tests, and simple analytical solutions
of this type exist in piecewise exponential potentials. Finally, the climbing
phenomenon can naturally inject slow--roll inflation: this is true even in the
one--field model motivated by \cite{sugimoto}, provided one takes into account
the stable non--BPS D3 brane that was identified in \cite{dms} following
\cite{sen}. There is clearly a delicate point, however. The climbing phase
occurs near the Big Bang, when higher-derivative corrections ($\alpha'$
corrections, in string language) are in principle large. Truly enough, they
could be small if the typical scale of the scalar potential were much smaller
than the string scale, but this is certainly not the case for the model of
\cite{sugimoto}. In lower dimensions, a similar problem presents itself for
string--size internal spaces, and in particular in the examples discussed in
Section \ref{sec:kklt}, where trapping is more natural precisely for small
{v.e.v.}'s of the moduli. We do not have concrete answers to all these
questions, but the climbing phenomenon seems nonetheless a suggestive, natural
and interesting option for the Early Universe in String Theory, with a
potential signature in the low--frequency tail of the CMB spectrum.


\scs{A Climbing Scalar in $D$ Dimensions}\label{sec:ten}


Let us consider a class of low--energy effective actions of the type
\be
S \ = \ {1 \over {2\kappa^2}} \, \int d^{D} x \, \sqrt{-g}\,
    \left[ \, R \, - \, {1 \over 2}\ (\partial \phi)^2
            \, - \, V(\phi) \, + \, \ldots \right]\, ,
\ee
in $D$ dimensions and with generic potentials $V(\phi)$. One can study an interesting class of
their cosmological solutions letting
\be
ds^2 \, =\, - \, e^{\,
2B(t)}\, dt^2 \, + \, e^{\, 2A(t)} \, d{\bf x} \cdot d{\bf x} \ ,
\qquad
\phi = \phi(t) \ , \label{metric}
\ee
and making the convenient gauge choice \cite{halliwell,dm2,townsend}
\be
V(\phi) \, e^{2B} = M^2 \, , \label{gauge}
\ee
where $M$ is a mass scale related to the potential $V(\phi)$. Working in terms of the ``parametric'' time $t$, that eq.~\eqref{metric} relates to the actual cosmological time $\eta$ according to
\be
d \eta \, = \, e^B \, dt \ ,\label{parametric}
\ee
and making the further redefinitions
\be \beta \, = \,
\sqrt{\frac{D-1}{D-2}}\ , \qquad \tau \,=\, M\, \beta \, t \ , \qquad \vf = \frac{\beta \, \phi}{\sqrt{\,2}} \ , \qquad a \,=\, (D-1)\, A\
, \label{defs} \ee
in an expanding universe, where
\be \dot{a} = \sqrt{1 \, +\, \dot{\vf}^2} \ , \ee
one is thus led to
\be\label{eqphi} \ddot{\vf} \, + \, \dot{\vf} \, \sqrt{1\,+\, \dot{\vf}^{\,2}}
\, +\, \left(\, 1+ \dot{\vf}^{\,2}\,\right)\ \frac{1}{2V}\ \frac{\pr V}{\pr
\vf}\, \, =\, 0\, , \ee
where ``dots'' denote $\tau$--derivatives.

\scss{Exponential potentials and exact solutions}

For the class of exponential potentials
\be V \ = \ M^2 \ e^{\,2\, \gamma \, \vf} \label{pot10} \ee
eq.~\eqref{eqphi} reduces to
\be\label{eqphi2}
\ddot{\vf} \, + \, \dot{\vf} \, \sqrt{1\,+\, \dot{\vf}^{\,2}} \, + \, \gamma \,
\left(\, 1+ \dot{\vf}^{\,2}\,\right)\, \, =\, 0\,
,
\ee
and up to a reflection one can restrict the attention to positive values of
$\g$. In general one can solve rather simply eq.~\eqref{eqphi2} letting ${\dot
\vf} = \sinh f $, and in particular in the ``critical'' $\gamma=1$ case
\be \label{sol10} \vf \, = \, \vf_0 +  \frac{1}{2} \
\log\left|\tau-\tau_0\right| \,-\, \frac{ \left(\tau-\tau_0\right)^{\,2}}{4} \
, \qquad a \, = \, a_0 +  \frac{1}{2} \ \log\left|\tau-\tau_0\right| \,+ \,
\frac{ \left(\tau-\tau_0\right)^{\,2}}{4} \, . \ee

A closer look reveals an amusing property of this solution: $\tau_0$ merely
defines the Big Bang, while the other integration constants fix the values of
the two functions $\vf(\tau)$ and $a(\tau)$ at some later reference time. As a
result, rather remarkably, $\vf$ \emph{can only emerge from the Big Bang while
climbing up the potential}. The scalar field then reverts its motion at
$\tau^*=\tau_0+1$, giving rise to a couple of $e$--folds of accelerated
expansion before the final descent along the potential. Letting $\tau_0=0$, for
all positive values of $\tau$ and $\g=1$ one thus finds
\be  ds^2 \ = \ e^{\frac{2a_0}{D-1}} \ |\tau|^{\frac{1}{D-1}} \
e^{\frac{\tau^2}{2(D-1)}}  \ d{\bf x} \cdot d{\bf x} \ - \ e^{-\, 2\,
\vf_{\,0}}\ |\tau|^{\, -\, 1} \ e^{\frac{\tau^2}{2}}\ \left({{d \tau} \over M
\beta}\right)^2 \, , \ \ e^{\vf} \ = \ e^{\vf_0 }\ |\tau|^{\frac{1}{2}} \
e^{-\frac{\tau^2}{4}} \, . \ee

For small values of $\gamma$ there should be no preferred motion for the scalar
field, and indeed for $\gamma<1$ eq.~\eqref{eqphi2} does admit two types of
solutions. The first describes again a scalar that emerges from the Big Bang
while climbing up the potential but, in its eventual descent, approaches from
below, in the ``parametric'' time $\tau$, the \emph{finite} limiting speed
\be\label{vlim}
v_l \,=\, - \
\frac{\gamma}{\sqrt{1-\gamma^{\,2}}} \ .
\ee
On the other hand, for $\g<1$ the second solution describes a scalar that
emerges from the Big Bang while climbing down the potential, at a speed in
``parametric'' time that eventually approaches from above the limiting value
\eqref{vlim}, but it disappears altogether as $\g \to 1$. However, the
suggestive analogy with eqs.~\eqref{mechanical} holds only insofar as one
refers to the ``parametric'' time $\tau$, or equivalently to $t$, since in all
cases the scalar comes eventually to rest in terms of the cosmological time.
Keeping this in mind, the complete solutions for $\g<1$ are
\be \begin{split} & ds^2 = e^{\frac{2a_0}{D-1}} \,
  \,\left|\sinh\left(\frac{\tau}{2}\, \sqrt{1-\g^2} \right)\right|^{\frac{2}{(1+\g)(D-1)}} \
\!\!\! \left[\cosh\left(\frac{\tau}{2}\, \sqrt{1-\g^2} \right)\right]^{\frac{2}{(1-\g)(D-1)}}\ d{\bf x} \cdot d{\bf x}
\\ &- e^{- \,2 \,\g \,\vf_0} \,
\left|\sinh\left(\frac{\tau}{2}\, \sqrt{1-\g^2} \right)\right|^{-\,\frac{2\g}{1+\g}}
\ \!\!\! \left[\cosh\left(\frac{\tau}{2}\, \sqrt{1-\g^2} \right)\right]^{\frac{2\g}{1-\g}}\,
\left({{d \tau} \over M \beta}\right)^2 \ ,  \\
&  e^{\vf} \ = \ e^{\vf_0 }\,
\left[\sinh\left(\frac{\tau}{2}\, \sqrt{1-\g^2} \right)\right]^{\frac{1}{1+\g}}
\left[\cosh\left(\frac{\tau}{2}\, \sqrt{1-\g^2} \right)\right]^{-\, \frac{1}{1-\g}}\,
\, \label{sgu1}
\end{split}
\ee
for the \emph{climbing} scalar, and
\be \begin{split} & ds^2 = e^{\frac{2a_0}{D-1}} \,
  \,\left|\cosh\left(\frac{\tau}{2}\, \sqrt{1-\g^2} \right)\right|^{\frac{2}{(1+\g)(D-1)}} \
\!\!\! \left[\sinh\left(\frac{\tau}{2}\, \sqrt{1-\g^2} \right)\right]^{\frac{2}{(1-\g)(D-1)}}\ d{\bf x} \cdot d{\bf x}
\\ &- e^{- \,2 \,\g \,\vf_0} \,
\left|\cosh\left(\frac{\tau}{2}\, \sqrt{1-\g^2} \right)\right|^{-\,\frac{2\g}{1+\g}}
\ \!\!\! \left[\sinh\left(\frac{\tau}{2}\, \sqrt{1-\g^2} \right)\right]^{\frac{2\g}{1-\g}}\,
\left({{d \tau} \over M \beta}\right)^2 \ ,  \\
&  e^{\vf} \ = \ e^{\vf_0 }\,
\left[\cosh\left(\frac{\tau}{2}\, \sqrt{1-\g^2} \right)\right]^{\frac{1}{1+\g}}
\left[\sinh\left(\frac{\tau}{2}\, \sqrt{1-\g^2} \right)\right]^{-\, \frac{1}{1-\g}}\,
\, \label{sgu2}
\end{split}
\ee for the \emph{descending} scalar. As we anticipated, the large--$\tau$
behavior of eqs.~\eqref{sgu1} and \eqref{sgu2} is the same: it corresponds to
the ``attractor solution'' of Lucchin and Matarrese \cite{attractor}, which can
be obtained from the previous expressions replacing $\tau$ with $\tau-\tau_0$
and considering the formal limit $\tau_0\to -\infty$. This behavior guarantees
that, at slow roll, a system of this type give rise to power--like inflation
\cite{attractor,pli}. In Section \ref{sec:climbing_inflation} we shall briefly
retrace how this can only occur for $\g< 1/\sqrt{D-1}$, well below the
``critical value'' for the climbing behavior, so that this simple one--field
model cannot combine climbing with steady--state inflation.

There is also a ``supercritical'' region of parameter space, which is
characterized by logarithmic slopes $\g > 1$. In this case there are two
singularities at the ends of the \emph{finite} interval $\tau \in
(0,{\pi}/{\sqrt{\g^2-1}})$ of ``parametric'' time, which spans the whole
cosmological evolution. The scalar continues to emerge from the Big Bang while
climbing up the potential, experiences a turning point as in the previous cases
and then climbs down the potential, approaching an infinite speed in
``parametric'' time but still coming to rest in terms of the cosmological time
$\eta$. The corresponding expressions for the space--time metric and the string
coupling,
\be \begin{split} & ds^2 = e^{\frac{2a_0}{D-1}} \,
  \,\left[\sin\left(\frac{\tau}{2}\, \sqrt{\g^{\,2}-1} \right)\right]^{\frac{2}{(1+\g)(D-1)}} \
\!\!\! \left[\cos\left(\frac{\tau}{2}\, \sqrt{\g^{\,2}-1} \right)\right]^{\,-\, \frac{2}{(\g-1)(D-1)}}\ d{\bf x} \cdot d{\bf x}
\\ &- e^{- \,2 \,\g \,\vf_0} \,
\left[\sin\left(\frac{\tau}{2}\, \sqrt{\g^{\,2}-1} \right)\right]^{-\,\frac{2\g}{1+\g}}
\ \!\!\! \left[\cos\left(\frac{\tau}{2}\, \sqrt{\g^{\,2}-1} \right)\right]^{\,-\, \frac{2\g}{\g-1}}\,
\left({{d \tau} \over M \beta}\right)^2 \ ,  \\
&  e^{\vf} \ = \ e^{\vf_0 }\,
\left[\sin\left(\frac{\tau}{2}\, \sqrt{\g^{\,2}-1} \right)\right]^{\frac{1}{1+\g}}
\left[\cos\left(\frac{\tau}{2}\, \sqrt{\g^{\,2}-1} \right)\right]^{\, \frac{1}{\g-1}}\,
, \label{supercrit}
\end{split}
\ee
can be obtained from those of eqs.~(\ref{sgu1}) and (\ref{sgu2}) as analytic
continuations.

Let us stress that the climbing solutions afford in principle a perturbative
realization in String Theory. On the contrary, the descending solutions reach
inevitably into regions of strong coupling at early times.  Finally, the
asymptotic behavior for large cosmological time $\eta$ of the scale factor is
$a \sim \eta^\frac{1}{\g^2(D-1)}$ for both the climbing and descending
solutions available for $\g<1$, in compliance with the Lucchin--Matarrese
attractor \cite{attractor}, and is simply $a \sim \eta^\frac{1}{(D-1)}$ for $\g
\geq 1$.

\scss{String Realizations}\label{sec:lower}

The phenomenon that we have illustrated plays a role in String Theory in at least two different contexts. The first is ``brane supersymmetry breaking'', in particular with reference to the model of \cite{sugimoto}, whose potential is induced from Riemann surfaces of Euler number one taking into account the $\overline{D9}$--$O9_+$ system present in the vacuum. The corresponding Einstein frame action is
\be
S_{10} = {1 \over {2\kappa_{10}^2}} \ \int d^{10} x \sqrt{-g}
\  \left\{  \left[ R \, - \, {1 \over 2}\, (\partial \phi)^2
             \, - \, 2 \, \a \, e^{3 \phi \over 2} \right] - \frac{1}{12} \ e^{\phi} \, H^2  \right\} \ , \label{st1}
\ee
where $H = d C + \omega_3$ is the field--strength of the RR two--form $C$ and
$\a=32\, T_9$ is the dilaton tadpole in units of the elementary D9 brane
tension. A similar expression would obtain after a toroidal compactification to
D space--time dimensions, but with
\be
\g \, =\, \frac{D+2}{\sqrt{2\, (D-1)\, (D-2)}} \ ,
\ee
so that this type of system is always ``supercritical'' for $D < 10$ and ``subcritical'' for $D > 10$, the additional range available for bosonic strings. The ten--dimensional model of \cite{sugimoto} sits in the middle, and corresponds precisely to the ``critical'' case $\g=1$. The second context is provided by non--critical strings, where the exponential potential arises on the sphere, and retracing the previous steps one could see that in this case $\g>1$ for $D\leq 4$ and $\g<1$ for $D\geq 5$.

One can now compactify \eqref{st1} to four dimensions, letting \cite{witten}
\begin{equation}
g_{i{\bar j}}^{(10)} \ = \ e^{\sigma} \delta_{i {\bar j}} \ , \quad g_{\mu
 \nu}^{(10)} \ = \ e^{-3 \sigma} g_{\mu \nu}^{(4)} \ \ ,
\end{equation}
where, in the absence of the tadpole, $i,{\bar j}=1,2,3$ would label the
complex coordinates of a Calabi-Yau space with $(h_{(1,1)}, h_{(2,1)}) =
(1,0)$. Aside from the breathing mode $\sigma$ and the dilaton $\phi$, two
axion fields would then behave as flat directions in four dimensions. While the
tadpoles are somehow treated as a perturbation, these suggestive steps have the
virtue of leading rather directly to the KKLT setting of \cite{kklt}. Indeed,
one can now define
\begin{equation}
 s \ = \ e^{3 \sigma}  e^{\phi \over 2} \  = \ e^{\Phi_s} \ , \qquad t \ = \ e^{
 \sigma}  e^{-{\phi \over 2}} \ = \ e^{{1 \over \sqrt 3} \ \Phi_t} \ ,
\end{equation}
where $\Phi_{s}$ and $\Phi_t$ are canonically normalized four--dimensional
fields, and working with $\kappa_4=1$ the relevant four--dimensional Lagrangian
reads
\begin{equation}
S_4 \ = \ {1 \over {2}} \int d^4x \, \sqrt{-g}\, \left[ R \,- \, {1 \over 2} \, (\partial
\Phi_{s})^2 \,- \, {1 \over 2} \, (\partial
\Phi_{t})^2 \, -\, 2\,
\alpha_1 \ e ^{-\sqrt{3}
 \Phi_t} + \cdots \, \right] \, \label{gamma4} .
\end{equation}
For the model of \cite{sugimoto} $s$ defines a flat direction, and therefore we
shall confine our attention to the solution $s=s_0$, that in principle may be
stabilized adding fluxes as proposed in \cite{kklt}. Once this is done, the
redefinitions \eqref{defs} show that the four--dimensional exponential
potential for $\Phi_t$ has precisely $\gamma=1$. One can thus say that the
ten--dimensional model of \cite{sugimoto} remains critical after this
compactification.

Another noteworthy option, a potential that is the sum of two exponentials, one
steep enough to induce the climbing and another flat enough to support a
slow--roll inflationary phase, is also available in the setup of
\cite{sugimoto}. In fact, the ten--dimensional Sugimoto model admits a
\emph{stable} non--BPS D3 brane \cite{dms}, in whose presence the complete
four--dimensional potential,
\begin{equation}
V \ = \  2 \alpha_1 \ e^{-\sqrt{3} \Phi_t} + \ \alpha_2 \ e^{- \frac{3
\Phi_s}{2} - \frac{\sqrt{3} \Phi_t }{2}}  \ , \label{nonbps3}
\end{equation}
becomes precisely of this type if $\Phi_s$ is somehow stabilized.

\scs{Moduli stabilization, climbing and trapping}\label{sec:kklt}

In the last few years, important progress in the study of string
compactifications with fluxes \cite{gkp} has triggered an intense activity on
the issue of moduli stabilization. The potentials of an interesting class of
models of this type were introduced in the classic KKLT paper \cite{kklt}. It
is thus interesting to investigate the behavior of these systems from our
vantage point.

\scss{Climbing in the KKLT system}

Let us consider a four--dimensional effective action described via a superpotential $W$ and a K\"ahler potential $K$ of the type
\be
W \ = \ W_0 \ + \ a \ e^{- b T} \ , \qquad K \ = \ - \, 3 \ \ln (T + {\bar T}) \ ,
\ee
where we work again with $\kappa_4 = 1$. In the standard setting of \cite{cfgv} these determine the potential
\begin{equation}
V_F \ = \ \frac{b}{(T + {\bar T})^2} \ \left\{ a \, {\bar W}_0 e^{- b T} \, + \, {\bar a} \, W_0 e^{- b {\bar T}}
\, + \, \frac{|a|^2}{3} \ [6 + b (T + {\bar T})] \ e^{- b (T + {\bar T})} \ \right\}
\ , \label{kklt1}
\end{equation}
and this class of models has tiny wells whose local minima correspond to
negative values for the vacuum energy. In order to overcome this problem, the
complete KKLT potentials of \cite{kklt} contain an additional contribution of
the type
\be
V \ = \ V_F \ + \ \frac{c}{(T + {\bar T})^3} \ , \label{uplift}
\ee
whose net effect is precisely to lift the minima to \emph{positive} vacuum
energies. This contribution is usually ascribed to an $F$--term uplift
\cite{uplift}, but as we have seen our arguments of Section \ref{sec:lower} can
relate it to a ten--dimensional tadpole. The complete potential has a valley of
local minima and maintains a typical runaway behavior in the asymptotic region
$Re \ T \rightarrow \infty$, where it is dominated by the uplift
\eqref{uplift}\footnote{In \cite{kklt}, a different uplift generated by a
$\overline{D3}$ anti--brane tension in the presence of warping led to a
potential $V \sim 1/t^2$. In our language, this contribution would correspond
to a ``subcritical'' logarithmic slope.}.

In adapting eqs.~\eqref{kklt1} and \eqref{uplift} to the four--dimensional KKLT system \cite{kklt}, the complex field $T$ is to be expanded according to
\begin{equation}
T \ =  \ e^{\Phi_t \over \sqrt{3} } \ + \ i \ \frac{\theta}{\sqrt{3}} \ , \label{kklt3}
\end{equation}
in terms of the canonically normalized scalar $\Phi_t$ and the axion $\theta$.
As we have anticipated, the last term in eq.~\eqref{uplift} corresponds
precisely to the ``critical'' value $\gamma = 1$, in the notation of Section
\ref{sec:ten}, so that the relevant portion of the low--energy effective field
theory reads
\begin{equation}
S = \frac{1}{2} \int d^4 x \sqrt{-g}
\left[ R - \frac{1}{2}\ (\partial \Phi_t)^2 - \frac{1}{2}\ e^{- \, \frac{2}{\sqrt{3}}\,
    \Phi_t} \ (\partial \theta)^2  - V (\Phi_T,\theta)
\right] \ . \label{kklt5}
\end{equation}
In the convenient gauge \eqref{gauge} and with the redefinitions
\begin{equation}
\Phi_t \, =\, \frac{2}{\sqrt{3}} \ x \ , \quad \theta \, = \, \frac{2}{\sqrt{3}} \ y \ , \quad \tau \, = \, M \, \sqrt{\frac{3}{2}} \ t \ ,
\label{kklt07}
\end{equation}
where $M$ is a dimensionful quantity related to the energy scale of the
potential $V$, and neglecting the contribution of the D9 brane (the D3 brane,
in the notation of the previous section), the field equations become
\begin{eqnarray}
\frac{d^2 x}{d\tau^2}&+& \frac{dx}{d\tau}\ \sqrt{1 +
  \left(\frac{dx}{d\tau}\right)^2 + e^{- \frac{4x}{3}}\,
  \left(\frac{dy}{d\tau}\right)^2} \, +\, \frac{1}{2\, V}\
\frac{\partial V}{\partial x}\ \left[1 +
  \left(\frac{dx}{d\tau}\right)^2 \right]\nonumber \\ &+&
 \frac{1}{2\, V}\ \frac{\partial V}{\partial y} \frac{dx}{d\tau} \frac{dy}{d\tau} \, +\, \frac{2}{3}\ e^{- \frac{4x}{3}} \left(\frac{dy}{d\tau}\right)^2 \, = \, 0  \ , \ \nonumber \\
\frac{d^2 y}{d\tau^2}&+& \frac{dy}{d\tau} \ \sqrt{1 +
  \left(\frac{dx}{d\tau}\right)^2 + e^{- \frac{4x}{3}}\,
  \left(\frac{dy}{d\tau}\right)^2} \, +\, \left( \frac{1}{2\, V}\
  \frac{\partial V}{\partial x} - \frac{4}{3} \right) \
\frac{dx}{d\tau}\ \frac{dy}{d\tau} \nonumber \\ &+& \frac{1}{2\,V}\
\frac{\partial V}{\partial y} \ \left[ e^{\frac{4x}{3}} + \left(\frac{dy}{d\tau}\right)^2\right] \ \, = \, 0 \ ,
\label{kklt8}
\end{eqnarray}
while the scalar potential takes the form
\begin{eqnarray}
V &=& \frac{c}{8} \ e^{-2 x} \,+\, \frac{b}{2} \ e^{-{4 x \over 3} - b \ e^{ 2 x \over 3}}
\left[ (Re \ a \overline{W_0}) \, \cos{2 b y \over 3} \,+\,
(Im \ a \overline{W_0})\, \sin{2 b y \over 3}  \right.  \nonumber \\
 &+& \left.  \frac{|a|^2}{3}\ \left(3 + b \ e^{ 2 x \over 3}\right) \ e^{-  b \ e^{ 2 x \over 3}}  \right] \ .
\label{kklt9}
\end{eqnarray}

Let us now focus on the ``critical'' tail of this potential, leaving aside
momentarily the tiny well and neglecting the contribution of the non--BPS D
brane. It is convenient to work in a slightly more general context, letting
\be
\frac{1}{2V}\ \frac{\partial V}{\partial x} \ = \ - \ \gamma\ , \qquad \frac{1}{2V}\ \frac{\partial V}{\partial y} \ = \ 0 \ ,
\ee
where $\gamma$ is actually 1 for the KKLT model. In this case $x$ and $y$ enter
eqs.~\eqref{kklt8} only via their derivatives, and our experience with the
one--field model of Section \ref{sec:ten} suggests the additional change of
variables
\be
\frac{dx}{d\tau}\, = \, r \, w \ , \qquad e^{\, - \, \frac{2x}{3}}\ \frac{dy}{d\tau}\, = \, r \, \sqrt{1-w^2} \ ,
\ee
with $w \in [-1,1]$, that finally reduces the system \eqref{kklt8} to
\begin{eqnarray}
&& \frac{dr}{d\tau} \,+ \ r\, \sqrt{1+r^2} \ - \, \gamma \, w \, \left(1+r^2\right) \, = \, 0 \ , \nonumber \\
&& \frac{dw}{d\tau}\, +\, (1-w^2)\ \left( \frac{2}{3} \ r \, - \, \frac{\g}{r}\right) \, = \, 0 \, . \label{redkklt}
\end{eqnarray}
The first equation is now strikingly similar to eq.~\eqref{eqphi2}, up to the
redefinition $r \to - \, \dot{\vf}$. The key novelty, that as we shall see
shortly has a remarkable effect on the dynamics, is that the parameter $\gamma$
of Section \ref{sec:ten} is replaced by $\g \, w$, that can assume any value in
the interval $[-\g,\g]$. As a result, this class of models can in principle
\emph{combine} the existence of a \emph{stable attractor} with the
\emph{climbing behavior} of Section \ref{sec:ten}. This is indeed the case, as
we now come to explain.

Let us begin by displaying attractor solutions for the non--linear system
\eqref{redkklt}. The first, more conventional one, is a $\tau$--independent
solution that can be found almost by inspection, and there are actually two
solutions of this type. One, with $w(\tau)=\pm1$, is again the
Lucchin--Matarrese attractor \cite{attractor} of the one--field model, while
the other,
\be r(\tau) \, = \,  \sqrt{\frac{3 \, \g}{2}} \ , \quad w(\tau) \, = \,
\frac{1}{\sqrt{\g\, \left(\g \,+ \, \frac{2}{3}\right)}}\ , \label{attractor2}
\ee
involves in an essential way both $\Phi_t$ and $\theta$ and exists provided
\be
\g \, \geq \, \frac{\sqrt{10}-1}{3} \ \approx 0.72 \ , \label{region}
\ee
so that it is available in the actual KKLT system, for which as we have seen
$\g = 1$. Below this value, the large--$\tau$ behavior of the system is
dominated by a different \emph{asymptotic} attractor that we originally noticed
in numerical tests, whereby
\be
\frac{dx}{d\tau} \, \sim \, c \ , \qquad y \, \sim \, e^{\frac{2 x}{3}} \ \alpha \ e^{-k \tau} \ ,
\label{lowerattractor}
\ee
with two constants $c$ and $\a$ and where $k$ must be non negative in order
that these contributions be bounded as $\tau \to \infty$. Interestingly, the
second of eqs.~\eqref{kklt8} determines $k$, and for the two-field system there
is thus a new option,
\be
 c \,= \, \frac{\gamma}{\sqrt{1-\gamma^2}} \quad , \quad
k \, = \, \frac{1}{\sqrt{1-\gamma^2}}\ \left[ 1 - \gamma \left(\gamma \, +\,
\frac{2}{3} \right) \right] \, , \ee
where $k > 0$ within a range of values for $\gamma$ that is complementary to
that of eq.~\eqref{region}. The attractors \eqref{attractor2} and
\eqref{lowerattractor} are stable in the corresponding ranges for $\g$, and in
particular for the system with the ``uplift'' \eqref{uplift}
eqs.~\eqref{attractor2} imply the typical large--$\tau$ behavior
\be
 \Phi_t(\tau) \, \sim \, \sqrt{\frac{6}{5}} \ \tau  \quad , \quad
 \theta(\tau) \, \sim \, \frac{1}{\sqrt{2}} \ \exp\left(\,\sqrt{\frac{8}{5}}\ \tau \,\right) \ .
 \label{attractorphitheta}
\ee

The system \eqref{redkklt} has an apparent singularity at $r=0$, but one can
show that the scalar simply reverts its motion before reaching this special
point. On the other hand, the large--$r$ behavior is particularly interesting
for our purposes, since it is typical of epochs that are close to the Big Bang.
The scalar moves very fast in this case, in terms of both $\tau$ and
cosmological time, so that the actual KKLT system (with $\g=1$) reduces to
\be
\frac{dr}{d\tau} \,+ \, \left(\e \,-\, w \right) r^2 \, \approx \, 0 \ , \quad
 \frac{dw}{d\tau}\, +\, \frac{2}{3} \ r (1-w^2) \, \approx \, 0 \ , \label{redfastkklt}
\ee
where $\e$ denotes the sign of $r$. These two equations can be combined into a single second--order equation for $r$ alone that integrates simply to
\be
\dot{r} \, \approx \, r^{\frac{8}{3}}\ C \, - \, 2\, \e \, r^2 \, , \label{newfirstfast}
\ee
where $C$ is a constant, but in this fashion one introduces spurious solutions
of eqs.~\eqref{redfastkklt} unless $C$ vanishes. As a result,
eq.~\eqref{newfirstfast} gives finally
\be r \, \approx \, \frac{1}{2\, \e \, \tau} \ , \label{new_2_firstfast} \ee
and the first of eqs.~\eqref{redfastkklt} then forces $w$ to approach $- \ \e \
$ as $|r|$ grows. Once $w$ gets frozen in this fashion, it should not come as a
surprise to the reader that one is led back to the one--field behavior, and in
fact combining this result with eq.~\eqref{new_2_firstfast} finally implies
that
\be
\frac{dx}{d \tau} \, \approx \, - \ \frac{1}{2\,\tau} \ ,
\ee
which describes indeed a climbing scalar.

In conclusion, as in the simpler one--field model of Section \ref{sec:ten} the
scalar field $\Phi_t$ is forced to emerge from the Big Bang while climbing up
the $\g=1$ potential, but in this case it eventually converges on the attractor
\eqref{attractorphitheta}. This typical behavior is seen very nicely in
numerical solutions of the full KKLT system.

\scss{Piecewise exponentials and trapping}
%
\begin{figure}[h]
\epsfxsize=0.27\textwidth
\centerline{\epsfbox{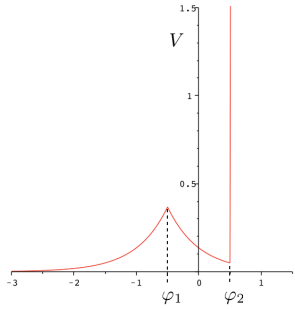}}
\caption{Piecewise-exponential  potentials lead to simple analytic trapping solutions.}
\end{figure}
It is intuitively clear that a climbing scalar can get trapped inside a
potential well if it can overcome the barrier and cosmological friction damps
its subsequent motion to a sufficient extent. As a result, the climbing
phenomenon can give rise to a variety of potentially interesting metastable
states. One can solve exactly eq.~\eqref{eqphi} for the instructive class of
``piecewise-exponential'' potentials, which can model a variety of potential
wells and thus open an instructive window on this phenomenon. The analytic
solutions can be obtained solving eq.~\eqref{eqphi2} in the various exponential
regions, as in Section \ref{sec:ten}, and then patching the results together by
demanding that $\vf$ and its first $\tau$--derivative be continuous at the
transition points where $\g$ changes abruptly. The reader will not fail to
notice the analogies with standard techniques used for the Schr\"odinger
equation in square--well potentials.

Let us illustrate the procedure for the class of potentials displayed in fig.~1,
\be
V \, = \,
\left\{
\begin{array}{ll}
M^2 \, e^{2 \vf} &
 \mbox{if \ \ \ } \vf < \vf_1  \ \ \mbox{\quad\qquad ( Region \ I )} \ ,
\\
M^2 \, e^{4\vf_1} \, e^{-\,2 \vf} &
 \mbox{if \ \ \ } \vf_1 \leq \vf < \vf_2 \mbox{\ \ \, ( Regions \ II , \ III )} \ ,
\\
\infty & \mbox{if \ \ \ } \vf \geq \vf_2  \ ,
\end{array}
\right.
\ee
where on the right we are actually introducing an infinite wall, which suffices to
illustrate the phenomenon and leads to simpler solutions of the matching conditions. To
this end, let us consider a scalar field that emerges from the Big Bang while climbing up
the outer wall of fig.~1, and for later convenience let us define the function
\be f(z) \, = \, {1 \over 2}\, \ln z \, -\, {z^2 \over 4} \ , \label{functf} \ee
so that, if the Big Bang occurs at $\tau=0$, in Region I
\be
 {\dot \vf}_{\rm I} \, = \, {1 \over {2\tau}} \, -\, {1 \over 2}\ \tau \
, \qquad
 \vf_{\rm I} \, = \, \vf^{(0)} \ +\ f(\tau) \ .
\ee
In order to enter the well, the scalar field must now reach the top of the barrier while
climbing up, and this is possible provided
\be \vf_1 \, - \, \vf^{(0)} \, \equiv \, f(\tau_1) \, < \, - \, \frac{1}{4} \ , \qquad 0
\, < \, \tau_1 \, < \, 1 \ . \label{regionI} \ee
In a similar fashion, the solution in Region II includes two integration constants,
$\tau^{(1)}$ and $\vf^{(1)}$, and reads
\be
 {\dot \vf}_{\rm II} \, =\, - \, {1 \over {2(\tau - \tau^{(1)})}}
                      \, + \, {1 \over 2} \, (\tau - \tau^{(1)}) \ , \qquad \vf_{\rm II} \, = \, \vf^{(1)} \, - f(\tau - \tau^{(1)})
   \ .
\label{regionII} \ee
Finally, the third region coincides with the second, that the scalar $\vf$ retraces after
being reflected by the infinite wall, so that $\vf_{III}$ takes again the form
\eqref{regionII}, albeit with two different integration constants $\tau^{(2)}$ and
$\vf^{(2)}$:
\be
 {\dot \vf}_{\rm III} \, = \, - \, {1 \over {2(\tau - \tau^{(2)})}}
                      \, + \, {1 \over 2} \, (\tau - \tau^{(2)}) \ , \qquad
 \vf_{\rm III} \, = \, \vf^{(2)} \, - \, f(\tau - \tau^{(2)}) \ .
\label{regionIII} \ee
\begin{figure}[h]
\epsfxsize=0.27\textwidth \centerline{\epsfbox{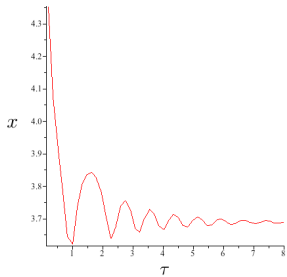}} \caption{A trapped
solution in a KKLT potential.}
\end{figure}

The matching conditions require that $\vf$ and its first derivative be
continuous at the ``parametric'' time $\tau_1$ when the top of the barrier is
first reached, so that
\be \vf_1 \, = \, \vf_I(\tau_1)\, =\, \vf_{II}(\tau_1) \ , \qquad  \dot{\vf}_I(\tau_1)\,
=\, \dot{\vf}_{II}(\tau_1) \ , \label{condphi1} \ee
and that a reflection occur at the ``parametric'' time $\tau_2$ when $\vf$
reaches the infinite wall:
\be \vf_2 \, = \, \vf_{II}(\tau_2)\, =\, \vf_{III}(\tau_2) \ , \qquad
\dot{\vf}_{II}(\tau_2)\, =\, - \, \dot{\vf}_{III}(\tau_2) \ . \label{condphi2} \ee
The conditions on the derivatives in eqs.~\eqref{condphi1} and \eqref{condphi2} are then
simple to solve, and give
\be \x \ \equiv \ \tau_1 \ =\ \frac{1}{\tau_1- \tau^{(1)}} \ , \qquad  \h \ \equiv \
\tau_2 - \tau^{(1)} \ = \ \frac{1}{\tau_2 - \tau^{(2)}}\ , \label{matchder} \ee
while trapping occurs if
\be \vf^{(2)} \, + \, \frac{1}{4} \ > \ \vf_1 \ ,  \ee
\emph{i.e.} if in region $III$ the scalar field reverts its motion before
reaching again $\vf_1$. In terms of the function $f$ of eq.~\eqref{functf} this
condition reads
\be f\left(\frac{1}{\x}\right) \, + \, f\left(\frac{1}{\h}\right) \, -\, f(\h) \, + \,
\frac{1}{4} \, > \, 0 \,  , \label{trapping_boundary} \ee
and implies that
\be \Delta \vf \, \equiv \, \vf_2 \, - \, \vf_1 \, > \, - \ \frac{1}{4} \ - \
f\left(\frac{1}{\h}\right) \ . \ee
In the range of interest matters simplify to a large extent, since $0<\x<1$ in
order that the scalar be climbing the outer wall when it reaches $\vf_1$ (for a
``fast'' scalar actually $\x<<1$). Then, on account of eq.~\eqref{matchder},
$\tau_1 - \tau^{(1)}>1$, and thus a fortiori $\h > 1$. As a result,  the
boundary of the trapping region \eqref{trapping_boundary} is well approximated
by the hyperbola $\x\, \h \, = \, 1$ (and particularly for a ``fast'' scalar),
so that one can finally conclude that trapping does occur in this model
provided
\be \Delta \vf  \, > \, - \ \frac{1}{4} \ - \ f(\x) \, = \, - \ \frac{1}{4} \, +\,
|\vf_1\, -\, \vf^{(0)}| \label{bound} \ . \ee
Notice that this is a rather weak condition, in view of the logarithmic growth
of $\Delta \vf$ with the ``speed'' $\dot{\vf}(\tau_1)$ of the scalar field at
$\vf_1$, the top of the outer barrier of the potential well.

Beyond this class of examples, there is some concrete evidence that trapping
occurs for wide ranges of parameters in the presence of ``critical'' or
``overcritical'' exponential potentials. For example, fig.~2 displays a
numerical solution of this type in a KKLT potential.

\scs{Inflation driven by climbing scalars}
\label{sec:climbing_inflation}
%
\begin{figure}[h]
\epsfxsize=0.27\textwidth \centerline{\epsfbox{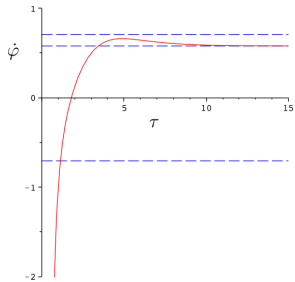}} \caption{Climbing
and inflation for the one--field system with the potential \eqref{nonbps3}.
Inflation occurs within the strip $\left|\dot{\vf}\right| < 1/\sqrt{2}$, while
the lower horizontal line in the upper portion of the plot is the attractor
determined by the D3-brane potential, $\dot{\vf}=1/\sqrt{3}$.}
\end{figure}
For the class of metrics \eqref{metric} that we have analyzed, the acceleration of the Universe is to be defined with reference to the cosmological time $\eta$, and thus occurs if
\begin{equation}
{\cal I} \, = \, \frac{d^{\, 2} A}{dt^{\, 2}} \ + \ \frac{d A}{dt} \left( \frac{d A}{dt}  \ - \ \frac{d B}{dt} \right)
\label{pl1}
\end{equation}
is \emph{positive}. In an expanding Universe, the acceleration can be
quantified via the corresponding number ${\cal N}$ of $e$--folds, where
\be \frac{d {\cal N}}{dt} \, = \, \frac{1}{\frac{d A}{dt}}\ {\cal I} \, = \,
\frac{d}{dt} \left[ \, \log \left(\frac{dA}{dt}\right) \ + \ A  \ - \ B \,
\right]  \  . \label{pl2} \ee
For the one--scalar system of Section \ref{sec:ten}, one can thus see that an
accelerated phase is possible if
\be \left|\dot{\vf}\right| \, < \, \frac{1}{\sqrt{D-2}} \ , \ee
and for instance for the Lucchin--Matarrese attractor this is the case only if
\be
\g \, < \, \frac{1}{\sqrt{D-1}} \ , \label{slowroll}
\ee
which lies well below $\g=1$, the ``critical'' logarithmic slope for the
climbing phenomenon. In a similar fashion, one can show that in the two--field
KKLT system the condition for an accelerated expansion is simply, in the
notation of Section \ref{sec:kklt},
\be r \ < \ \frac{1}{\sqrt{2}} \ , \label{boundadot} \ee
which is not fulfilled during the final descent for $\g=1$. As we have
anticipated, however, the combined effects of the D9--brane tadpole of
\cite{sugimoto} and of the non--BPS D3--brane tadpole of \cite{dm2} can lead to
a climbing phase that is eventually followed by steady--state inflation, since
the first term of eq.~\eqref{nonbps3} forces the scalar field to climb up when
emerging from the Big Bang while the second dominates the eventual descent
where it can indeed support slow--roll inflation. Fig.~3 displays a numerical
solution of this type.

\section*{Acknowledgments}
We are grateful to P.~Creminelli, D.~Langlois, C.~Papineau, S.~Patil,
S.~Pokorski, V.~Rubakov, K.~Turszynski, and especially to E.~Kiritsis, for
useful discussions, and to APC--Paris VII, CPhT--Ecole Polytechnique and Scuola
Normale Superiore for the kind hospitality extended to one or more of us. This
work was supported in part by the contract PITN-GA-2009-237920, by the ERC
Advanced Grants n. 226371 (MassTeV) and n. 226455 ``Supersymmetry, Quantum
Gravity and Gauge Fields'' (SUPERFIELDS), by the CNRS PICS no. 3747 and 4172,
by Scuola Normale Superiore, by INFN, by the MIUR-PRIN contract 2007-5ATT78,
and by the Grant-in-Aid for Scientific Research No.19540303 from the Ministry
of Education, Culture, Sports, Science and Technology of Japan.


\begin{thebibliography}{99}

\bibitem{cosmology}
For recent reviews see:\\
V.~Mukhanov, ``Physical foundations of Cosmology,'' {\it  Cambridge, UK: Univ.
Pr. (2005)}; \\ S.~Weinberg, ``Cosmology,'' {\it Oxford, UK: Oxford Univ. Pr.
(2008)}.

\bibitem{weinberg_ft}
For a recent review, see:\\
S.~Weinberg, ``The Quantum Theory of Fields, 3 vols.,'' {\it Cambridge, UK:
Cambridge Univ. Pr. (2000)}.

\bibitem{fs}
W.~Fischler and L.~Susskind,
 Phys.\ Lett.\  B {\bf 171}, 383 (1986),
 Phys.\ Lett.\  B {\bf 173}, 262 (1986).

\bibitem{wv}
E.~Dudas, M.~Nicolosi,G.~Pradisi and A.~Sagnotti,
 Nucl.\ Phys.\  B {\bf 708} (2005) 3
 [arXiv:hep-th/0410101];
N.~Kitazawa,
 Phys.\ Lett.\  B {\bf 660} (2008) 415
 [arXiv:0801.1702 [hep-th]].

\bibitem{orientifolds}
A.~Sagnotti, in Cargese '87, ``Non-Perturbative Quantum Field
Theory'', eds. G. Mack et al (Pergamon Press, 1988), p. 521,
arXiv:hep-th/0208020;
G.~Pradisi and A.~Sagnotti,
Phys.\ Lett.\ B {\bf 216} (1989) 59;
P.~Horava,
Nucl.\ Phys.\ B {\bf 327} (1989) 461,
Phys.\ Lett.\ B {\bf 231} (1989) 251;
M.~Bianchi and A.~Sagnotti,
Phys.\ Lett.\ B {\bf 247} (1990) 517,
Nucl.\ Phys.\ B {\bf 361} (1991) 519;
M.~Bianchi, G.~Pradisi and A.~Sagnotti,
Nucl.\ Phys.\ B {\bf 376} (1992) 365;
A.~Sagnotti,
 Phys.\ Lett.\  B {\bf 294}, 196 (1992)
 [arXiv:hep-th/9210127]~.
\\
For reviews see: E.~Dudas,
Class.\ Quant.\ Grav.\  {\bf 17}, (2000) R41 [arXiv:hep-ph/0006190];
C.~Angelantonj and A.~Sagnotti,
Phys.\ Rept.\  {\bf 371} (2002) 1 [Erratum-ibid.\  {\bf 376} (2003)
339] [arXiv:hep-th/0204089];
R.~Blumenhagen, B.~Kors, D.~Lust and S.~Stieberger,
 Phys.\ Rept.\  {\bf 445} (2007) 1
 [arXiv:hep-th/0610327].

\bibitem{sugimoto} S.~Sugimoto,
Prog.\ Theor.\ Phys.\  {\bf 102} (1999) 685 [arXiv:hep-th/9905159].

\bibitem{bsb}
I.~Antoniadis, E.~Dudas and A.~Sagnotti,
Phys.\ Lett.\ B {\bf 464} (1999) 38 [arXiv:hep-th/9908023];
C.~Angelantonj,
Nucl.\ Phys.\ B {\bf 566} (2000) 126 [arXiv:hep-th/9908064];
G.~Aldazabal and A.~M.~Uranga,
JHEP {\bf 9910} (1999) 024 [arXiv:hep-th/9908072];
C.~Angelantonj, I.~Antoniadis, G.~D'Appollonio, E.~Dudas and
A.~Sagnotti,
Nucl.\ Phys.\ B {\bf 572} (2000) 36 [arXiv:hep-th/9911081].


\bibitem{nlsusy}
D.~V.~Volkov and V.~P.~Akulov,
 Phys.\ Lett.\  B {\bf 46} (1973) 109 .

\bibitem{dm1}
E.~Dudas and J.~Mourad,
 Phys.\ Lett.\  B {\bf 514} (2001) 173
 [arXiv:hep-th/0012071];
G.~Pradisi and F.~Riccioni,
 Nucl.\ Phys.\  B {\bf 615}, 33 (2001)
 [arXiv:hep-th/0107090].

\bibitem{dm2}
 E.~Dudas and J.~Mourad,
 Phys.\ Lett.\  B {\bf 486} (2000) 172
 [arXiv:hep-th/0004165].

\bibitem{townsend}
P.~K.~Townsend and M.~N.~R.~Wohlfarth,
  Phys.\ Rev.\ Lett.\  {\bf 91} (2003) 061302
  [arXiv:hep-th/0303097];
 Class.\ Quant.\ Grav.\  {\bf 21} (2004) 5375
  [arXiv:hep-th/0404241];
R.~Emparan and J.~Garriga,
  JHEP {\bf 0305} (2003) 028
  [arXiv:hep-th/0304124];
J.~G.~Russo,
  Phys.\ Lett.\  B {\bf 600} (2004) 185
  [arXiv:hep-th/0403010].

\bibitem{augustopheno}
A. Sagnotti, talk at ``String Phenomenology 2010'' (Paris, July 6\ 2010).

\bibitem{attractor}
F.~Lucchin and S.~Matarrese,
 Phys.\ Rev.\  D {\bf 32} (1985) 1316.

\bibitem{halliwell}
J.~J.~Halliwell,
  Phys.\ Lett.\  B {\bf 185} (1987) 341.

\bibitem{pli}
L.~F.~Abbott and M.~B.~Wise,
 Nucl.\ Phys.\  B {\bf 244} (1984) 541;
D.~H.~Lyth and E.~D.~Stewart,
 Phys.\ Lett.\  B {\bf 274} (1992) 168.

\bibitem{exponential}
I.~P.~C.~Heard and D.~Wands,
  Class.\ Quant.\ Grav.\  {\bf 19} (2002) 5435
  [arXiv:gr-qc/0206085];
N.~Ohta,
  Phys.\ Rev.\ Lett.\  {\bf 91} (2003) 061303
  [arXiv:hep-th/0303238];
S.~Roy,
  Phys.\ Lett.\  B {\bf 567} (2003) 322
  [arXiv:hep-th/0304084].

\bibitem{bergshoeff}
E.~Bergshoeff, A.~Collinucci, U.~Gran, M.~Nielsen and D.~Roest,
  Class.\ Quant.\ Grav.\  {\bf 21} (2004) 1947
  [arXiv:hep-th/0312102];
A.~Collinucci, M.~Nielsen and T.~Van Riet,
  Class.\ Quant.\ Grav.\  {\bf 22} (2005) 1269
  [arXiv:hep-th/0407047].

\bibitem{kklt}
S.~Kachru, R.~Kallosh, A.~Linde and S.~P.~Trivedi,
  Phys.\ Rev.\  D {\bf 68} (2003) 046005
  [arXiv:hep-th/0301240];
S.~Kachru, R.~Kallosh, A.~D.~Linde, J.~M.~Maldacena, L.~P.~McAllister and S.~P.~Trivedi,
  JCAP {\bf 0310} (2003) 013
  [arXiv:hep-th/0308055].

\bibitem{dms}
 E.~Dudas, J.~Mourad and A.~Sagnotti,
 Nucl.\ Phys.\  B {\bf 620} (2002) 109
 [arXiv:hep-th/0107081].

\bibitem{sen}
 A.~Sen,
  JHEP {\bf 9808} (1998) 010
  [arXiv:hep-th/9805019].

\bibitem{witten}
 E.~Witten,
 Phys.\ Lett.\  B {\bf 155} (1985) 151.

 \bibitem{noscale}
  E.~Cremmer, S.~Ferrara, C.~Kounnas and D.~V.~Nanopoulos,
  Phys.\ Lett.\  B {\bf 133} (1983) 61.

\bibitem{gkp}
S.~B.~Giddings, S.~Kachru and J.~Polchinski,
  Phys.\ Rev.\  D {\bf 66} (2002) 106006
  [arXiv:hep-th/0105097].

\bibitem{uplift}
M.~Gomez-Reino and C.~A.~Scrucca,
  JHEP {\bf 0605} (2006) 015
  [arXiv:hep-th/0602246];
O.~Lebedev, H.~P.~Nilles and M.~Ratz,
  Phys.\ Lett.\  B {\bf 636} (2006) 126
  [arXiv:hep-th/0603047];
E.~Dudas, C.~Papineau and S.~Pokorski,
  JHEP {\bf 0702} (2007) 028
  [arXiv:hep-th/0610297];
 H.~Abe, T.~Higaki, T.~Kobayashi and Y.~Omura,
  Phys.\ Rev.\  D {\bf 75} (2007) 025019
  [arXiv:hep-th/0611024].

\bibitem{cfgv}
E.~Cremmer, S.~Ferrara, L.~Girardello and A.~Van Proeyen,
  Nucl.\ Phys.\  B {\bf 212} (1983) 413.
\end{thebibliography}
\end{document}